\documentclass[epj]{svjour}
\usepackage{graphics}
\usepackage{amsmath}
\begin{document}
\newcommand{\Ima}{\textrm{Im}}
\newcommand{\Rea}{\textrm{Re}}
\newcommand{\mev}{\textrm{ MeV}}
\newcommand{\gev}{\textrm{ GeV}}
\newcommand{\dtres}{d^{\hspace{0.1mm} 3}\hspace{-0.5mm}}

\newcommand{\eps}{\epsilon}
\newcommand{\half}{\frac{1}{2}}
\newcommand{\thalf}{\textstyle \frac{1}{2}}

\newcommand{\Nmass}{M_{N}} 
\newcommand{\delmass}{M_{\Delta}} 
\newcommand{\pimass}{\mu}  
\newcommand{\rhomass}{m_\rho} 
\newcommand{\piNN}{f}      
\newcommand{\rhocoup}{g_\rho} 
\newcommand{\fpi}{f_\pi} 
\newcommand{\f}{f} 
\newcommand{\nucfld}{\psi_N} 
\newcommand{\delfld}{\psi_\Delta} 
\newcommand{\fpiNN}{f_{\pi N N}} 
\newcommand{\fpiND}{f_{\pi N \Delta}} 
\newcommand{\GMquark}{G^M_{(q)}} 
\newcommand{\vecpi}{\vec \pi}
\newcommand{\vectau}{\vec \tau}
\newcommand{\vecrho}{\vec \rho}
\newcommand{\delmu}{\partial_\mu}
\newcommand{\delMu}{\partial^\mu}
\newcommand{\nn}{\nonumber}
\newcommand{\bi}{\bibitem}
\newcommand{\vs}{\vspace{-0.20cm}}
\newcommand{\be}{\begin{equation}}
\newcommand{\ee}{\end{equation}}
\newcommand{\ba}{\begin{eqnarray}}
\newcommand{\ea}{\end{eqnarray}}
\newcommand{\ropi}{$\rho \rightarrow \pi^{0} \pi^{0}
\gamma$ }
\newcommand{\roeta}{$\rho \rightarrow \pi^{0} \eta
\gamma$ }
\newcommand{\omepi}{$\omega \rightarrow \pi^{0} \pi^{0}
\gamma$ }
\newcommand{\omeeta}{$\omega \rightarrow \pi^{0} \eta
\gamma$ }
\newcommand{\ul}{\underline}
\newcommand{\del}{\partial}
\newcommand{\rth}{\frac{1}{\sqrt{3}}}
\newcommand{\rsix}{\frac{1}{\sqrt{6}}}
\newcommand{\sq}{\sqrt}
\newcommand{\fr}{\frac}
\newcommand{\pr}{^\prime}
\newcommand{\ov}{\overline}
\newcommand{\Gm}{\Gamma}
\newcommand{\rw}{\rightarrow}
\newcommand{\rgl}{\rangle}
\newcommand{\De}{\Delta}
\newcommand{\Dp}{\Delta^+}
\newcommand{\Dm}{\Delta^-}
\newcommand{\Dz}{\Delta^0}
\newcommand{\Dpp}{\Delta^{++}}
\newcommand{\Sg}{\Sigma^*}
\newcommand{\Sp}{\Sigma^{*+}}
\newcommand{\Sm}{\Sigma^{*-}}
\newcommand{\Sz}{\Sigma^{*0}}
\newcommand{\X}{\Xi^*}
\newcommand{\Xm}{\Xi^{*-}}
\newcommand{\Xz}{\Xi^{*0}}
\newcommand{\Om}{\Omega}
\newcommand{\Omm}{\Omega^-}
\newcommand{\kp}{K^+}
\newcommand{\kz}{K^0}
\newcommand{\pip}{\pi^+}
\newcommand{\pim}{\pi^-}
\newcommand{\piz}{\pi^0}
\newcommand{\et}{\eta}
\newcommand{\kb}{\ov K}
\newcommand{\km}{K^-}
\newcommand{\kbz}{\ov K^0}
\newcommand{\ksb}{\ov {K^*}}
\def \ketv #1>{\mbox{$|{#1}\rangle$}}
\def \etal{{\it et al.}}
\newcommand{\rts}{ \sqrt s}

\def\tstrut{\vrule height2.5ex depth0pt width0pt} 
\def\jtstrut{\vrule height5ex depth0pt width0pt} 
\title{$\rho \rho N$ and $\rho \rho \Delta$ molecules with $J^P=5/2^{+}$ and $J^P=7/2^{+}$}

\author{Bao-Xi Sun$^{1,2}$, Hua-Xing Chen$^{1,3}$ and E. Oset$^{1}$
}                     
\institute{Departamento de F\'{\i}sica Te\'orica and IFIC, Centro Mixto Universidad de Valencia-CSIC, Institutos de Investigaci\'on de Paterna, Apartado
22085, 46071 Valencia, Spain,
\and
Institute of Theoretical Physics, College of Applied Sciences, Beijing University of Technology, Beijing 100124, China
\and
School of Physics and Nuclear Energy Engineering, Beihang University, Beijing 100191, China}
\date{Received: date / Revised version: date}
%
\abstract{
The $\rho\rho N$ and $\rho\rho\Delta$ three-body systems have been studied within the framework of the fixed center approximation of Faddeev equation. The $\rho\rho$ interaction in isospin $I=0$, spin $S=2$ is strongly attractive, and so are the $N \rho$, $\Delta \rho$ interactions. This leads to bound states of both $\rho\rho N$ and $\rho\rho\Delta$. We find peaks of the modulus squared of the scattering matrix around 2227 MeV for $\rho\rho N$, and 2372 MeV for $\rho\rho\Delta$.
Yet, the strength of the peak for the $\rho\rho N$ amplitude is much smaller than for $\rho\rho\Delta$, weakening the case for a $\rho\rho N$ bound state, or a dominant $\rho\rho N$ component.
A discussion is made on how these states can be searched for in present programs looking for multimeson final states in different reactions.
} 
\maketitle
\section{Introduction}
\label{intro}

While it has been always accepted that the meson baryon cloud on top
of three constituent quarks plays some role in baryon observables, like
form factors and static properties, it is becoming gradually more
broadly accepted that in some cases this meson baryon cloud can even
account for the largest part of the wave function, as far as one is
concerned with physical processes taking place at small or intermediate
energies. The use of effective theories that provide interaction
Lagrangians in terms of the observable degrees of freedom, mesons and
baryons~\cite{Weinberg:1978kz}, has allowed to study systems made out of mesons
and baryons, and in some cases the interaction is so strong that one
gets bound states or resonances of that system. The systematic use of
chiral Lagrangians~\cite{Weinberg:1978kz,gasser,ecker,ulfrep} and unitary techniques in coupled
channels has lead to the so called chiral unitary approach which has
been rather successful describing $J^P=1/2^-$ low lying baryonic
resonances~\cite{Mag05} or the low lying $J^P=3/2^-$ states~\cite{lutz,sarkar}. Extensions have
been done for the interaction of vector mesons with baryons, both for
the octet of baryons~\cite{angelsvec} as for the decuplet~\cite{baosourav}.

Systems of two mesons and one baryon have also started to be studied.
Faddeev calculations of these systems have found that the low lying
$J^P=1/2^+$ baryonic states, up to the Roper which is a very complicated
object~\cite{juelich}, can be reproduced as bound states or resonances
of two mesons and one baryon in coupled channels~\cite{albertouno,albertodos}.

States of larger spin have also been investigated, involving vector
mesons and eventually the decuplet of baryons. For instance in
Ref.~\cite{vijande} an explanation was given for the  $\Delta(5/2^-)(1930)$
as a $\rho \Delta$ bound state. States of large spin with positive
parity have also been investigated as bound states or resonances of two
mesons and one baryon. For instance in Ref.~\cite{xiepedro} a plausible
explanation was found for a $\Delta(1/2^{+})$ state around 1750 MeV, which
is highly problematic in quark models \cite{pedro}, in terms of the $\pi
\rho \Delta$ system.  It is interesting to pursue this search, looking
for states with higher spin which are also difficult to obtain with
quark models, and which can be easier accommodated by means of meson
baryon components. In this paper we go along this line and study systems
made out of two $\rho$ mesons and a $N$ or $\Delta$.

In order to evaluate the amplitudes for the interaction of these
three body systems we take advantage of the fact that the $\rho \rho$
interaction in $I=0$ and spin $S=2$ is huge and leads to the bound state of
the $f_2(1270)$ resonance, which decays into $\pi \pi$ as found in Ref.~\cite{raquel}.
The system is so bound and packed because of the short
range of the interaction that the presence of an extra baryon should not
alter very much this strongly bound $\rho \rho$ system. This is an ideal
situation to apply the Fixed Center Approximation (FCA) to the Faddeev
equations~\cite{Chand:1962ec,Toker:1981zh,Kamalov:2000iy,Gal:2006cw}, which certainly simplifies the technical
work. Since we are not interested in a precise determination of the
energies, and uncertainties of even 50 MeV are acceptable, the tool
should be more than sufficient to state the existence of bound states of
these systems and their approximate energy. We are reassured in this
task by recent works dealing with analogous three hadron systems, for
instance the $K \bar{K}N$ system in Ref.~\cite{Xie:2010ig}, where one could compare the results of the
FCA with those of more elaborate Faddeev calculations~\cite{albertodos}
or variational calculations~\cite{Jido:2008kp}. Also a study of the
$\bar{K}NN$ system in Ref.~\cite{melahat} leads to similar results as those
obtained with variational methods using the same elementary amplitude
for the $\bar{K} N$ interaction~\cite{Dote:2008in}.

The fact that we start with the $\rho \rho $ system in $S=2$ allows us
to get bound states of the two systems studied with spins $S=3/2, 5/2$ for
the $\rho \rho N$ and $S=1/2, 3/2, 5/2, 7/2$ for  $\rho \rho \Delta$. Such high
spin states are difficult to obtain with the right energy in quark
models, which make the structure suggested in this work a more plausible
ansatz for the nature of these states.
This is the case of the two $\Delta(5/2^+)$ around 1740 MeV and 2200 MeV, which have been discussed in Ref.~\cite{xiepedro}.

\section{the interaction of vector meson and octet baryon}

In the Fixed Center Approximation (FCA), we shall need the $\rho N$ interaction and the form factor of the $f_2(1270)$ resonance. First we show the result of the $\rho N$ interaction we used, which has been obtained in Ref.~\cite{angelsvec}. We use the Bethe-Salpeter equation in coupled channels ($\rho N$, $\omega N$, $\phi N$, $K^* \Lambda$ and $K^* \Sigma$), and the scattering matrix is given in matrix form as
\begin{equation}
t = V+V G t = [1-VG]^{-1}V,
\end{equation}
where the kernel $V$ and the meson-baryon two-body propagator $G$ are also given in Ref.~\cite{angelsvec}. In dimensional regularization the propagator $G$ is given by
\begin{eqnarray}
\label{eq:g-function}
&& \mbox G_i(s, m_i, M_i)
\\ \nonumber &=&  \frac{2 M_i}{(4 \pi)^2}
  \Biggr \{
        a_i(\mu) + \log \frac{m_i^2}{\mu^2} +
        \frac{M_i^2 - m_i^2 + s}{2s} \log \frac{M_i^2}{m_i^2}
  \\ \nonumber
    &+& \frac{Q_i(\rts)}{\rts}
    \left[
         \log \left(  s-(M_i^2-m_i^2) + 2 \rts Q_i(\rts) \right) \right.
  \\ && ~~~~~~~~~~ +  \log \left(  s+(M_i^2-m_i^2) + 2 \rts Q_i(\rts) \right)
  \nonumber
  \\ \nonumber
  && ~~~~~~~~~~~
  \Biggl.
      - \log \left( -s+(M_i^2-m_i^2) + 2 \rts Q_i(\rts) \right)
   \\
  && ~~~~~~~~~~ \left.  - \log \left( -s-(M_i^2-m_i^2) + 2 \rts Q_i(\rts) \right)
    \right]
  \Biggr\},
  \nonumber
\end{eqnarray}
which is identical to that in Ref.~\cite{angelsvec}, although written in a different form. The magnitudes $m_i$ and $M_i$ denote the masses of the two propagating particles. In the present case they are vector mesons and octet baryons, respectively. The parameters $\mu$ and $a_i$ are fitted with the experimental data of vector meson-baryon scattering. Here we shall use the values $\mu=630$ MeV and $a_i(\mu)=-2.0$, which have been found in Ref.~\cite{angelsvec} to lead to a good description of a few vector baryon resonances (note that $\mu$ and $a_i(\mu)$ are not independent parameters and there is only one degree of freedom).

Under the approximation of neglecting the three-momentum of the vector meson versus its mass, as done in Ref.~\cite{angelsvec} and followed here, the potential $V$ is given by
\be
V_{ij}=-\frac{1}{4f_\pi^2}C_{ij}(k^0+k^{\pr 0})\vec{\epsilon}\cdot\vec{\epsilon}^\prime,
\label{poten}
\ee
where $f_\pi=93MeV$ is the pion decay constant, $\vec{\epsilon}$, $\vec{\epsilon}^\prime$ the polarization vectors of the initial and final vector and $k^0$, $k^{\pr 0}$ the energies of the initial and final vector mesons. The coefficients $C_{ij}$ are given in Ref.~\cite{angelsvec} for the different transitions between the coupled channels. This potential stems from the exchange of a vector meson between the external vectors and the baryon, involving a three vector vertex which is provided by the local hidden gauge Lagrangian \cite{hidden1}.

For the $\rho \rho$ system we do not need the interaction but only the wave function. It is obtained using the interaction of Ref.~\cite{hidden1} with a four-vector contact term and diagrams with vector exchange in the $t$ and $u$ channel involving two three-vector vertices \cite{raquel}, constructing the wave function following the approach of Ref.~\cite{YamagataSekihara:2010pj}.

Since the vector mesons, particularly the $\rho$ and the $K^*$, are rather
broad, one has to take into account their widths. To do this, we follow Ref.~\cite{angelsvec}
and replace the $G$ function appearing in Eq.~(\ref{eq:g-function}) by
\begin{eqnarray}\label{eq:rhowidth}
\tilde{G}(s,m,M) &=&\frac{1}{N_m}\int^{(m+2\Gamma_m)^2}_{(m-2\Gamma_m)^2}
d\tilde{m}^2 \times
\\ \nonumber &\times& \left(-\frac{1}{\pi}\right)~\mathrm{Im}\frac{1}{\tilde{m}^2-m^2+i m \tilde{\Gamma} (\tilde{m})}
G(s,\tilde{m},M) \, ,
\end{eqnarray}
where the normalization factor $N_m$ is
\begin{eqnarray}
N_m&=&\int^{(m+2\Gamma_m)^2}_{(m-2\Gamma_m)^2}
d\tilde{m}^2\left(-\frac{1}{\pi}\right)\mathrm{Im}\frac{1}{\tilde{m}^2-m^2+i m \tilde{\Gamma} (\tilde{m})} \, ,
\nonumber \\
\end{eqnarray}
and the ``width function'' $\tilde{\Gamma}(\tilde{m})$ is defined to be
\begin{equation}
\tilde{\Gamma}(\tilde{m})=\Gamma_0  \frac{m^2}{\tilde{m}^2}  \frac{q^3_\mathrm{off}}{q^3_\mathrm{on}}\Theta(\tilde{m}-m_1-m_2) \, ,
\end{equation}
where $\Gamma_0$ is the decay width of the vector meson, and $m_1$, $m_2$ are the masses of the two pseudoscalar mesons, to which the vector mesons decay. For the $\rho$ meson, they are $m_1=m_2=m_\pi$, and \begin{equation}
q_\mathrm{off}=\frac{\lambda^{1/2}(\tilde{m}^2,m_\pi^2,m_\pi^2)}{2\tilde{m}},\quad
q_\mathrm{on}=\frac{\lambda^{1/2}(M_\rho^2,m_\pi^2,m_\pi^2)}{2 M_\rho},
\end{equation}
For $K^*$ mesons, they are $m_1=m_\pi$ and $m_2=m_K$, and
\begin{equation}
q_\mathrm{off}=\frac{\lambda^{1/2}(\tilde{m}^2,m_K^2,m_\pi^2)}{2\tilde{m}},\quad
q_\mathrm{on}=\frac{\lambda^{1/2}(M_{K^*}^2,m_K^2,m_\pi^2)}{2 M_{K^*}},
\end{equation}
where $\lambda$ is the K\"allen function.

In principle, since one has coupled channels to the $\rho N$, $\omega N$, $\phi N$, $K^* \Lambda$ and $K^* \Sigma$, the three body system should contain $\rho \omega N$, $\rho \phi N$ etc. states, in particular the $\rho \omega N$ state is close in energy to the $\rho \rho N$ and could play a role. The strategy of considering all coupling channels to evaluate the most important amplitude and then neglect the less important channels in the three body system has been followed in similar systems, like the $\bar{K} N N$, where the $\bar{K} N$ amplitude is evaluated within the $\bar{K} N$ and $\pi \Sigma$ coupling channels, but the $\pi \Sigma N$ channels are ignored in the three body system (beyond what is implicitly accounted for in the $\bar{K} N$ amplitude) \cite{Dote:2008in}.
In Ref.~\cite{melahat} it was, however, shown that the explicit consideration of the $\pi \Sigma N$ channel in the three body system led to minor changes, once the effect of the $\pi \Sigma$ channel was incorporated in the $\bar{K} N$ amplitude. In the present case there are strong reasons to neglect the $\rho \omega N$, $\omega \omega N$ channels as explicit channels in the three body system. Indeed, the $\omega N$ amplitude at energies close to the $\rho N$ threshold is negligible \cite{angelsvec} and the $\omega \omega$ channel couples very weakly to $\rho \rho$ through the $f_2(1270)$ resonance \cite{geng} (the $\rho \omega$ channel appears in $I=1$ and is only relevant at energies around 1600 MeV~\cite{geng}).

\section{$N$-$f_2(1270)$ interaction}
\label{sec:nf2}

The essence of the FCA is that from the three body system one has reasons to consider that two particles are strongly bound making a cluster and the third particle interacts moderately with the components of the cluster such as not to modify its wave function. Then one ignores the coupled channel dynamics and considers the interaction of the third particle with the cluster, including all multiple scattering steps with the components of the cluster.

The formalism to evaluate the interaction for the $N\rho\rho$ system is similar to the one used in
Refs.~\cite{Roca:2010tf,YamagataSekihara:2010qk,xiepedro}, and we shall follow the same procedures and study the case
of $N-(\rho\rho)_{f_2(1270)}$, in particular \cite{xiepedro,YamagataSekihara:2010qk} where one deals with different particles. For the $N \rho \rho$ system, we will consider that
two of the $\rho$ mesons are clusterized forming an $f_2(1270)$ resonance, given the strong
binding of the $f_2(1270)$ system. This allows us to use the fixed center approximation (FCA) to the Faddeev equations.

\begin{figure}
\resizebox{0.4\textwidth}{!}{%
  \includegraphics{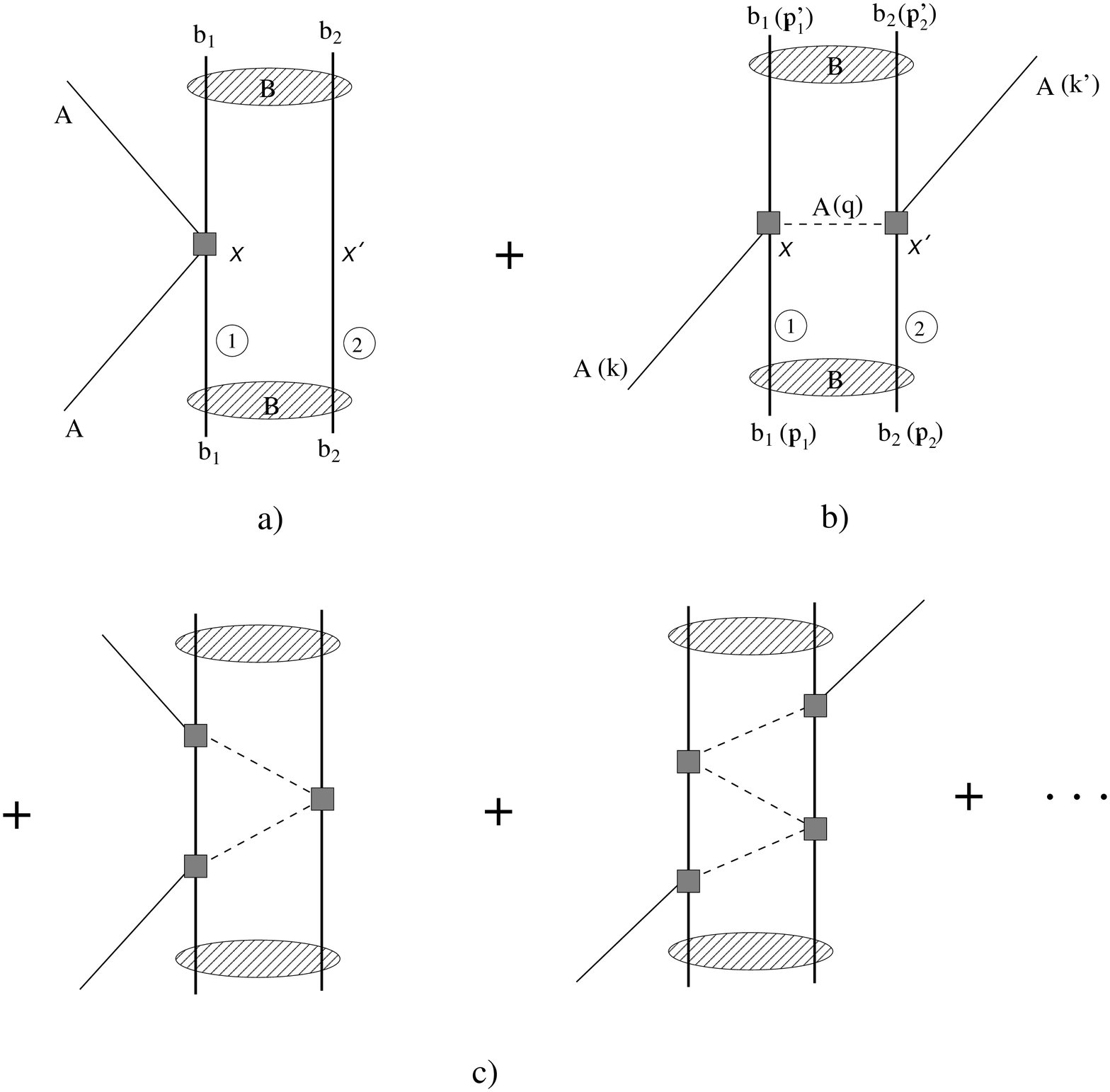}
}
\caption{Diagrammatic representation of the fixed center approximation
to the Faddeev equations for the interaction of a particle $A$ with a particle $B$
made of a cluster of two particles $b_1$ and $b_2$. Diagrams $a)$ and $b)$ represent
the single and double scattering contributions  respectively. In addition to these diagrams there are the equivalent ones where the external particle scatters first with the second particle in the clusters.}
\label{fig:Faddeev}
\end{figure}

The FCA to Faddeev equations is depicted diagrammatically in Fig.~\ref{fig:Faddeev}. The external particle, the nucleon in this case,
interacts successively with the other two $\rho$ mesons which form the
$\rho \rho$ cluster.  The FCA equations are written in terms of two
partition functions $T_1$ and $T_2$, which sum up to the total scattering
matrix
\ba
&&T_1= t_1+t_1 G_0 T_2 \nn \, , \\
&&T_2= t_2+t_2 G_0 T_1 \nn \, , \\
&&T_{N f_2(1270)} =T_1+T_2 \, ,
\ea
where $T_{N f_2(1270)}$ is the total scattering amplitude we are looking for; $T_i$
accounts for all the diagrams starting with the interaction
of the external particle with particle $i$ of the compound
system; $t_i$ represents the $N \rho$ unitarized scattering
amplitude of a proton with any of the $\rho$ in the $I=0$ $\rho\rho$ system; $G_0$ is the loop function for the particle N propagating
inside the compound system which will be discussed later on (see Eq.~(\ref{eq:G0})).
The schematic representation is
depicted in Fig.~\ref{fig:Faddeev}. Fig.~\ref{fig:Faddeev}a) represents the single-scattering contribution
and Fig.~\ref{fig:Faddeev}b) the double-scattering. The contributions of Fig.~\ref{fig:Faddeev}a) and b)
are the two first contributions of the Faddeev equations.

Since there are two $\rho$ mesons in the present case, we have
$T_1=T_2$, and thus we have the following equations:
\ba
&& T_1 = t_1+t_1 G_0 T_1 \nn \, , \\
&& T_{N f_2(1270)} = 2T_1 \, .
\label{eq:Faddeevred}
\ea

\subsection{Wave function of $f_2(1270)$}\label{sec:wave}

\begin{figure}
\resizebox{0.4\textwidth}{!}{%
  \includegraphics{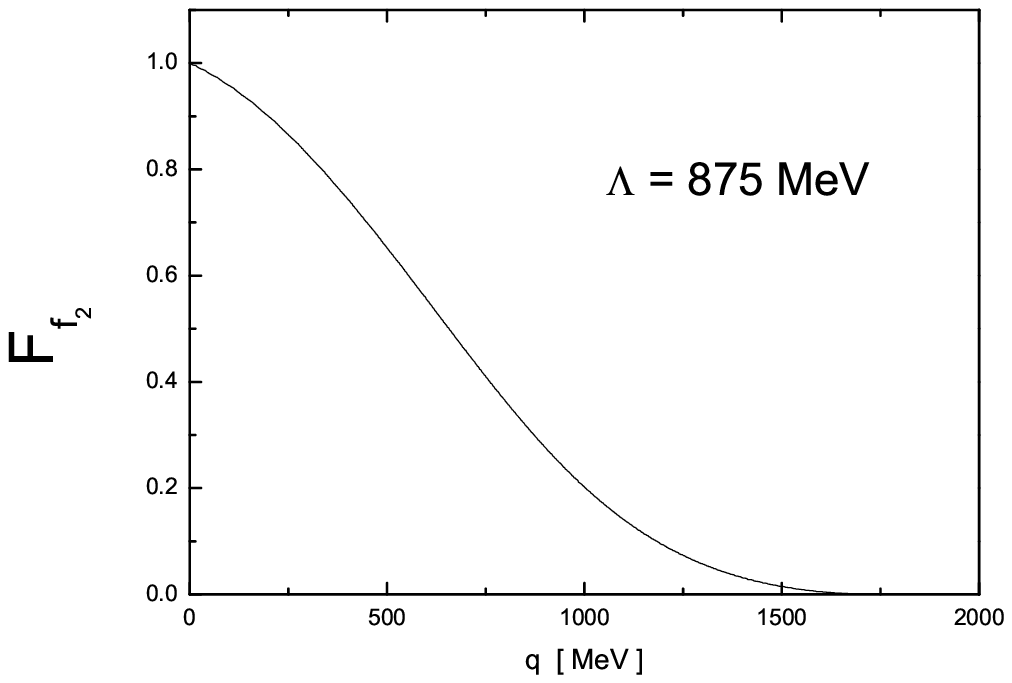 }
}
\caption{Form factor of the $f_2(1270)$ resonance with $\Lambda = 875 MeV$.}
\label{fig:Ff2}
\end{figure}

We assume that the two $\rho$ mesons are bound forming the ${f_2(1270)}$, whose
wave function and the corresponding form factor have been discussed
in Refs.~\cite{Roca:2010tf,YamagataSekihara:2010qk}. It has been shown that the use of a separable potential in momentum space of the type
\be
V=v\theta(\Lambda -q)\theta(\Lambda -q'),
\ee
where $\Lambda$ is the cutoff used in the theory for the scattering of
two $\rho$ mesons as in Ref.~\cite{raquel}, and $q,q'$ are the modulus of the momenta, leads to the
same on  shell prescription for the scattering matrix as is used in
the chiral unitary approach. This allows to calculate form factors consistently with the findings of the chiral unitary approach for the resonances which are dynamically generated~\cite{Gamermann,YamagataSekihara:2010pj}.
According to these references, we use the following form factor
\ba
\label{eq:form}
F_{f_2}(q)&=&\frac{1}{\cal N}
\int_{\substack{p<\Lambda\\|\vec p-\vec q|<\Lambda}}
d^3p\, \times
\\ &\times& \nonumber
\frac{1}{M_{f_2}-2\omega_\rho(\vec p)}\,
\frac{1}{M_{f_2}-2\omega_\rho(\vec p-\vec q)},
\label{eq:ff2cutoff}
\ea
where the normalization factor ${\cal N}$ is
\ba
{\cal N}=
\int_{p<\Lambda}d^3p
\frac{1}{\left( M_{f_2}-2\omega_\rho(\vec p) \right)^2}.
\ea
Here $M_{f_2}$ is the mass of the bound state ${f_2(1270)}$;
$\omega_\rho(\vec p)$ is the energy of the internal $\rho$ meson of
$f_2(1270)$. In Fig.~\ref{fig:Ff2} we show this form factor $F(q)$
as a function of $q$, where the cutoff parameter $\Lambda$ is chosen
to be 875 MeV.

\subsection{Single and Double Scattering Contribution}

We write the amplitude $t_1$ entering Eq.~(\ref{eq:Faddeevred})
in terms of the $I=3/2$ and $I=1/2$
unitarized amplitudes ($t_{N\rho}^{(I=3/2)}$, $t_{N\rho}^{(I=1/2)}$)
. For this we consider a cluster of two $\rho$ mesons in isospin $I=0$,
the constituents of which
we call mesons 1 and 2. The nucleon will be labeled with
number 3.
The two $\rho$ mesons forming the $f_2$ are in an $I=0$ state
\begin{eqnarray}
|\rho\rho\rangle_{I=0}&=&-\frac{1}{\sqrt{3}}
|\rho^+\rho^- + \rho^-\rho^+ + \rho^0\rho^0\rangle
\\ \nonumber &=& \frac{1}{\sqrt{3}}\Big( |(1,-1)\rangle
+ |(-1,1)\rangle - |(0,0)\rangle\Big)
\end{eqnarray}
where the kets in the last member of the equation indicate the $I_z$ components of the
1 and 2 particles, $|(I_z^{(1)},I_z^{(2)})\rangle$.
We take the nucleon in the state $|(I_z^{(3)})\rangle$
\be
|p\rangle=\bigg |( \frac{1}{2}) \bigg \rangle.
\ee

The scattering amplitude $t_1$ in terms of the two-body isospin amplitudes is given by
\ba
t_1+t_2 &=& \langle N (\rho \rho)_{f_2} | t(s) | N (\rho \rho)_{f_2} \rangle
\\ \nonumber &= &\left( \bigg\langle \frac{1}{2}\bigg| \otimes \frac{1}{\sqrt{3}} \left(\langle (+1,-1)+
 (-1,+1) -  (0,0)|\right)\right)
 \\ \nonumber && \times (t_{31} + t_{32}) \times \nn \\
& & \left(\bigg|\frac{1}{2}\bigg\rangle \otimes \frac{1}{\sqrt{3}} \left(|(+1,-1) +
 (-1,+1) - (0,0)\rangle\right) \right)\nn\\
\nonumber &=&\frac{1}{3}\left(2t_{31}^{I=3/2}(s_1)+t_{31}^{I=1/2}(s_1)\right)
\\ \nonumber &+& \frac{1}{3}\left(2t_{32}^{I=3/2}(s_2)+t_{32}^{I=1/2}(s_2)\right) \ea
where the notation followed in the last term for the states is
$\langle (I^{\textrm{total}} I_z^{\textrm{total}},
I_z^k)|t_{ij}|\rangle$, where $I^{\textrm{total}}$ means the total
isospin of the $ij$ system and $k\neq i,j$ (the spectator $\rho$).
This leads, in terms of the $I=3/2$ and $I=1/2$
unitarized amplitudes ($t_{N\rho}^{(I=3/2)}$, $t_{N\rho}^{(I=1/2)}$),
to the following amplitude for the single scattering contribution:
\be
\label{eq:t1-nrr}
t_1=\frac{1}{3}\left(2t_{N\rho}^{(I=3/2)}(s')+t_{N\rho}^{(I=1/2)}(s')\right),
\ee
where $s'$ is the invariant mass of the
external particle $N$ and the first/second particle inside the bound
state $f_2(1270)$:
\begin{equation}
s'=s_1=s_2=\frac{1}{2}\left( s+2m_\rho^2+M_N^2-M_{f_2}^2 \right) \, .
\end{equation}

Now the $S$-matrix for the single scattering term is given by
\ba
S_1^{(1)}&=&-it_1 \frac{1}{{\cal V}^2}
\frac{1}{\sqrt{2\omega_{p_1}}}
\frac{1}{\sqrt{2\omega_{p'_1}}}
\sqrt{\frac{M_N}{E_N(k)}}
\sqrt{\frac{M_N}{E_N(k')}} \times \nonumber
\\ && \times (2\pi)^4\,\delta(k+K_{f_2}-k'-K'_{f_2}),
\label{eq:Ssingle}
\ea
where ${\cal V}$ is the volume of the box where we normalize the states to unity.
The superindex of $S$ indicates single scattering and the subindex that the interaction occurs on particle 1 of the cluster.
 The final expression for the $S$-matrix for the double scattering process is
\ba
\label{eq:s2}
S_1^{(2)}&=&-i
\sqrt{\frac{M_N}{E_N(k)}}
\sqrt{\frac{M_N}{E_N(k')}}
\frac{1}{\sqrt{2\omega_{p_1}}}
\frac{1}{\sqrt{2\omega_{p'_1}}}
\frac{1}{\sqrt{2\omega_{p_2}}}
\frac{1}{\sqrt{2\omega_{p'_2}}}\nn\\ \nn
&&\times \int \frac{d^3q}{(2\pi)^3}
F_{f_2}(q)
\frac{M_N}{E_N(\vec{q})}\cdot
\frac{1}{{q^0}-E_N(\vec{q})+i\epsilon} t_1 t_1
\\ &&\times (2\pi)^4\delta(k+K_{f_2}-k'-K'_{f_2})\frac{1}{{\cal V}^2} .
\label{eq:finalS2}
\ea
The subindex indicates that the first interaction is in particle 1 of the cluster. In Eq.~(\ref{eq:s2}), we have also taken into account that $(\vec{k}+\vec{k'})/2=0$ on average.

\subsection{General form of the $S$-matrix}

The general form of $S$-matrix for the nucleon-$f_2(1270)$ interaction can be written as
\ba
\label{eq:S-matrix}
S_1&=&\delta_{fi}-iT_{N f_2}(s)\frac{1}{ {\cal V}^2 }
{\sqrt{\frac{M_N}{E(k)}}}
{\sqrt{\frac{M_N}{E(k')}}}
\frac{1}{\sqrt{2\omega_{f_2}}}
\frac{1}{\sqrt{2\omega_{{f_2}'}}} \nn
\\ && \times (2\pi)^4\,\delta(k+K_{f_2}-k'-K'_{f_2}),
\ea
where $i$ and $f$ indicate the initial and final states, respectively. By comparing the summation of $S^{(1)}$ and $S^{(2)}$ with the scattering matrix $S$ in Eq.~(\ref{eq:S-matrix}), the $N-f_2(1270)$ amplitude can be obtained as
\ba \nonumber
T_{N f_2}(s)&=&2\cdot\frac{\sqrt{2\omega_{f_2}}}{\sqrt{2\omega_{\rho}}}\frac{\sqrt{2\omega_{f'_2}}}{\sqrt{2\omega_{\rho'}}}
\Biggr(t_1+
\frac{1}{\sqrt{2\omega_{\rho}}}\frac{1}{\sqrt{2\omega_{\rho'}}}t_1\\ && \times\int\frac{d^3q}{(2\pi)^3}
F_{f_2}(q)
\frac{M_N}{E_N(\vec{q}~^{2})}\cdot
\frac{1}{{q^0}-E_N(\vec{q}~^{2})+i\epsilon}
 t_1 \Biggr) \nonumber \\
&=&2\left(t'_1+
t'_1 G_0(s)
 t'_1 \right),
 \label{eq:tnf2-fa2}
 \ea
 where
\begin{eqnarray}
t'_1  &=& t_1
\sqrt{\frac{2\omega_{f_2(1270)}}{2\omega_{\rho}}}
\sqrt{\frac{2\omega_{f_2(1270)}'}{2\omega_{\rho}'}}~\approx~t_1 \frac{M_{f_2}}{m_\rho}, \nonumber \\
G_0(s) &=& \frac{1}{\sqrt{2\omega_{f_2(1270)}
2\omega_{f_2(1270)}'}}  \int \frac{d^3\vec{q}}{(2\pi)^3}
\\ \nonumber && \times F_{f_2(1270)}(q)
\frac{M_N}{E_N(\vec{q})}\frac{1}{q^0-E_N(\vec{q}) +
i\epsilon} \nonumber \nonumber \\
&\approx& \frac{1}{2M_{f_2}}  \int \frac{d^3\vec{q}}{(2\pi)^3} F_{f_2(1270)}(q)
\frac{M_N}{E_N(\vec{q})}\frac{1}{q^0-E_N(\vec{q}) +
i\epsilon}, \nonumber
\label{eq:G0}
\end{eqnarray}
with $E_N(\vec{q})=\sqrt{\vec{q}~^{2}+M_N^2}$. In the center of mass system of the nucleon and the bound state of $f_2(1270)$, the energy $q^0$ carried by the nucleon is
$$q^0=\frac{s+M_N^2-M_{f_2}^2}{2\sqrt{s}}.$$

If the whole series of the expansion of $T_{N f_2}$ in the fixed center approximation is taken into account, the scattering amplitude $T_{N f_2}$  can be deduced from Eq.~(\ref{eq:Faddeevred}) and we have
\ba
T_{N f_2}(s)~=~\frac{2t'_1}{1-G_0 t'_1}~=~\frac{2}{t'^{-1}(s')_1-G_0(s) } \, .
\label{eq:tnf2prime}
\ea

\section{Consideration of the $f_2(1270)$ width}

The consideration of the $\rho \rho \rightarrow \pi \pi$ decay in $L=2$ in Ref.~\cite{raquel} gave rise to a width of the $f_2(1270)$, which has not
been taken into account in our previous analysis. To add its contribution, we modify the scattering matrix $T_{N f_2}(s)$ to
\begin{eqnarray}
 T_{N f_2}(s)&=&\frac{1}{N_{f_2}} \int_{{(m_{f_
 2}-2 \Gamma_{f_2})}^2}^{{(m_{f_
 2}+2 \Gamma_{f_2})}^2}
d \tilde{m}^2
\\ \nonumber & \times& \left( -\frac{1}{\pi} \right) \mbox{Im} \left[\frac{1}{\tilde{m}^2 - m^2_{f_2} +i \tilde{m} \Gamma_1(\tilde{m})}  \right]  T_{N f_2}(s, \tilde{m}),
\label{eq:massdistri}
\end{eqnarray}
where the normalization factor $N_{f_2}$ is
\begin{eqnarray}
N_{f_2} &=& \int_{{(m_{f_
 2}-2 \Gamma_{f_2})}^2}^{{(m_{f_
 2}+2 \Gamma_{f_2})}^2}
d \tilde{m}^2
\\ \nonumber &\times& \left( -\frac{1}{\pi} \right) \mbox{Im} \left[\frac{1}{\tilde{m}^2 - m^2_{f_2} +i \tilde{m} \Gamma_1(\tilde{m})}  \right],
\end{eqnarray}
and $\Gamma_1(\tilde{m})$ is proportional to the decay width of the $f_2$, $\Gamma_{f_2(1270)}=157$ MeV,
\begin{equation}
\Gamma_1(\tilde{m}) = \Gamma_{f_2(1270)}
\Big (\frac{\lambda^{1/2}(\tilde{m}^2,m_1^2,m_2^2)2 m_{f_2}}{\lambda^{1/2}(m_{f_2}^2,m_1^2,m_2^2)2\tilde{m}} \Big )^5, \end{equation}
where the exponent in the momenta is now $2L+1=5$.

\section{Extension to the case of the $\Delta \rho \rho$ system}

We can follow the same procedures to study the case of the $\Delta \rho \rho$ system. There are just a few points we need to pay attention to:
\begin{enumerate}

\item We need now the $\rho \Delta$ interaction. This one is obtained in a similar way to the $\rho N$ and is done in Ref.~\cite{baosourav}. There the coupled channels $\Delta \rho$ and $\Sigma^* K^*$ are taken into account, but the results are basically the same as those obtained using the single channel $\Delta \rho$ in Ref.~\cite{vijande}. Once again the interaction of $\Delta \rho \rightarrow \Delta \rho$ stems from $\rho$ exchange between the two external $\rho$ and the $\Delta$, involving the three vector vertex of the local hidden gauge approach \cite{hidden1}.

\item We need to take into account the decay widths of decuplet baryons. This is similar to what we have done in Eq.~(\ref{eq:rhowidth}).
To do this, we follow Ref.~\cite{angelsvec}, and replace the $\tilde{G}$-function appearing in Eq.~(\ref{eq:rhowidth}) by $\tilde{\tilde{G}}$:
\begin{eqnarray}
\tilde{\tilde{G}}(s,m,M)&=&\frac{1}{N_M}\int^{M+2\Gamma_M}_{M-2\Gamma_M}
d\tilde{M}
\\ \nonumber &\times& \left(-\frac{1}{\pi}\right)\mathrm{Im}\frac{1}{\tilde{M}-M+i{\tilde{\Gamma}' (\tilde{M})\over2}}
\tilde{G}(s,m,\tilde{M}) \, ,
\end{eqnarray}
where the normalization factor $N_M$ is
\begin{eqnarray}
N_M&=&\int^{M+2\Gamma_M}_{M-2\Gamma_M}
d\tilde{M}\left(-\frac{1}{\pi}\right)\mathrm{Im}\frac{1}{\tilde{M}-M+i{\tilde{\Gamma}' (\tilde{M})\over2}} \, , \nonumber \\
\end{eqnarray}
and the ``width function'' $\tilde{\Gamma}'(\tilde{M})$ is defined to be
\begin{equation}
\tilde{\Gamma}'(\tilde{M})=\Gamma_0 \Big ( { \lambda^{1/2}(\tilde{M}^2,M_1^2,M_2^2) 2 M \over \lambda^{1/2}(M^2,M_1^2,M_2^2) 2 \tilde{M} } \Big)^3\Theta(\tilde{M}-M_1-M_2) \, ,
\end{equation}
where $\Gamma_0$ is the decay width of the decuplet baryon; $M_1$ and $M_2$ are the masses of the octet baryon and pseudoscalar meson to which the decuplet baryon decays. Take the $\Delta$ baryon as an example, they are $M_1 = M_N$ and $M_2 = m_\pi$.

\item The amplitude for the single-scattering contribution is:
\be
t_1=\frac{1}{3}\left(3/2t_{\Delta \rho}^{I=5/2}+t_{\Delta \rho}^{I=3/2}+1/2 t_{\Delta \rho}^{I=1/2}\right).
\ee

\end{enumerate}
Besides this, we also need to change the relevant masses, etc. But all the procedures are similar and straightforward, and so we shall not discuss it further.

\section{Spin of the states}

As to the spin of systems, it is easy to determine. The cluster of $\rho \rho$ is the $f_2(1270)$ which has $J=S=2$. Then the nucleon interacts with the $\rho$ meson. At low energies, where the three-momentum of the $\rho$ meson can be neglected versus its mass, the $\rho N \rightarrow \rho N$ amplitude has the form $A \vec{\epsilon}\cdot \vec{\epsilon}^\prime$, with $\vec{\epsilon}$, $\vec{\epsilon}^\prime$ the initial and final $\rho$ polarization vectors \cite{angelsvec}. With no extra nucleon spin dependence and $L=0$, this interaction leads to degenerate $\rho N$ states with spin 1/2 and 3/2. For the case of the $\rho \Delta \rightarrow \rho \Delta$ interaction, the same argument holds, but now the degenerate $\rho \Delta$ states will have spins 1/2, 3/2 and 5/2. If we start from $\rho \rho$ in $S=2$, the final spin states of the three body system, which will appear degenerate in the approach will have $J^P=3/2^+, 5/2^+$ for the $\rho \rho N$ and $J^P=1/2^+, 3/2^+, 5/2^+, 7/2^+$ for the $\rho \rho \Delta$. The largest spin available would in principle be the most difficult to accommodate in quark models, and be better candidates for molecular states.

\section{Results}

\subsection{$N \rho \rho$ system}

\begin{figure}
\resizebox{0.4\textwidth}{!}{%
  \includegraphics{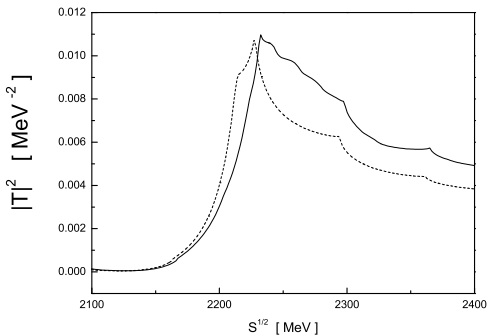}
}
\caption{Modulus squared of the unitarized $N-\rho-\rho$ amplitude with $\Lambda=875$ MeV. The solid line denotes the case with the $f_2(1270)$ decay width, and the dashed line is for the case without the $f_2(1270)$ decay width.}
\label{fig:tnf2}
\end{figure}

In Fig.~\ref{fig:tnf2}, we show the results of $|T|^2$ for the $N \rho \rho$ system with $\Lambda=875$ MeV, which is suited to obtain the $f_2(1270)$ resonance according to the work of Ref.~\cite{raquel}. We observe a peak around 2227 MeV with a width of 100 MeV. The peak does not have a standard Breit-Wigner form. The sharp peak could be indicative of a cusp effect but the threshold for $N f_2(1270)$ is at 2210 MeV, about 17 MeV below the peak in Fig.~\ref{fig:tnf2}. In fact, a small cusp peak at threshold is also visible in the figure at this energy. On the other hand, when the convolution for the mass distribution of the $f_2(1270)$ due to its width is considered, the peak of the cusp disappears, but a peak in $|T|^2$, slightly shifted to higher energies, still remains with a similar or slightly larger width.

The nonstandard shape of the peak, but particularly the small strength of $|T|^2$ compared to the $\rho \rho \Delta$ case (see Fig.~\ref{fig:tdf2}), should warn us about identification of this peak with a resonance. With such a small strength, other components of the wave function as $\rho K^* \Lambda$, etc., could play some role. We can only take this signal as a weak indication of a possible state around this energy in which the $\rho \rho N$ component might play a relevant role.

\subsection{$\Delta \rho \rho$ system}

\begin{figure}
\resizebox{0.4\textwidth}{!}{%
  \includegraphics{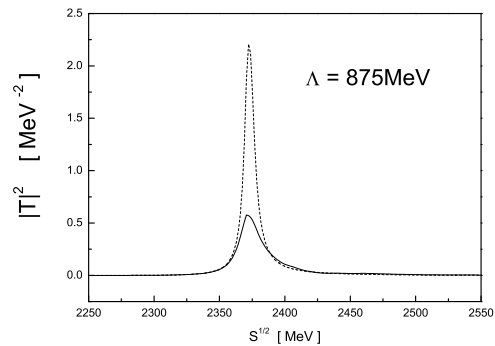}
}
\caption{Modulus squared of the unitarized $\Delta-\rho-\rho$ amplitude with $\Lambda=875$ MeV. The solid line denotes the case with the $f_2(1270)$ decay width, and the dashed line is for the case without the $f_2(1270)$ decay width.}
\label{fig:tdf2}
\end{figure}

\begin{figure}
\resizebox{0.4\textwidth}{!}{%
  \includegraphics{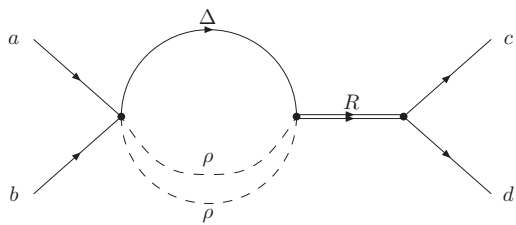}
}
\caption{Diagrammatic representation of the formation of a resonance R made of a bound $\Delta \rho \rho$ and decaying into particles c and d.}
\label{fig:product}
\end{figure}

In Fig.~\ref{fig:tdf2}, we show the results of $|T|^2$ for the $\Delta \rho \rho$ system with $\Lambda=875$ MeV. We observe a peak around 2372 MeV.
Two features can be observed in Fig.~\ref{fig:tdf2}. The peak has now an approximate Breit-Wigner shape and the strength of $|T|^2$ at the peak is about 200 times bigger than in Fig.~\ref{fig:tnf2} for the $N \rho \rho$ state, although the peak is now narrower than in Fig.~\ref{fig:tnf2}. Yet, the integrated strength of the peak is still about 40 times bigger. The large value of $T$ in the $\Delta \rho \rho$ case indicates that in a production of the resonance in one reaction, the magnitude of the resonance excitation would be large through the consideration of the intermediate $\Delta \rho \rho$ state and its coupling to the resonance (see Fig.~\ref{fig:product}).
The consideration of the width of the $f_2(1270)$ reduces the strength of the peak and increases the width of the resonance. The width is still relatively small, about 25 MeV.
In our approach, once the convolution for the width of the $f_2(1270)$ is done, the main decay channel would be $\Delta \pi \pi$. This is interesting to know from the experimental point of view.
This kind of information is useful since there is an increasing interest in looking at multimeson final states from the experimental point of view, and devoted programs are carried out at Hall D of Jefferson Lab \cite{pennington}, at COMPASS \cite{sebaneubert} and at LEPS in Spring8/Osaka, where $\Delta \rho$ final states are being investigated \cite{hicks}.

\section{Discussion and conclusions}

The region of energies around 2300-2400 MeV is poorly known concerning $N$ and $\Delta$ resonances. Although several resonances are compiled in the PDG \cite{pdg}, it seems clear that there is room for many more. Concerning the $N^*$ resonance obtained with $3/2^+$, $5/2^+$, there is no candidate in the states tabulated in the PDG. Our support for this state was also weak. On the other hand, the $\Delta$ state obtained with $1/2^+$, $3/2^+$, $5/2^+$, $7/2^+$ around 2372 MeV could find a counterpart in the $\Delta(2390)(7/2^+)$ \cite{pdg}, which has been observed in two independent experiments \cite{Cutkosky} with a mass of $2350\pm100$ MeV and in Ref.~\cite{hoehler} with a mass of $2425\pm60$ MeV. It is also worth mentioning that this state is not seen in the latest GWU analysis \cite{Arndt}.

In any case the state claimed in both \cite{Cutkosky,hoehler} has a width of around $300\pm100$ MeV, which seems not compatible with the narrow width of about 25 MeV of the state that we obtain. It is also worth recalling that masses of $N^*$ and $\Delta^*$ resonances in this energy range and with these quantum numbers can be reached with quark models \cite{Capstick}, hence maybe the narrow width is indicative of another kind of state, mostly made from a bound $\Delta \rho \rho$ as we have found. The two experiments done to claim the $\Delta(2390)(7/2^+)$ are done with analysis of the $\pi N \rightarrow \pi N$ reaction. As we have mentioned, the $\pi N$ is not the expected decay channel of the resonance found, but mostly $\Delta \pi \pi$. A different kind of analysis would have to be done to eventually find this resonance. The advent of programs looking for multimeson final states \cite{pennington,sebaneubert,hicks} offers unique opportunities to look for these states. The theoretical calculations, indicating the most favorable decay channels, can serve as a guideline for these experiments.

\section*{Acknowledgments}
We would like to thank Ju-Jun Xie for helpful discussions and Pedro Gonzalez for a careful reading and valuable information concerning quark models.
This work is partly supported by DGICYT contract  FIS2006-03438, the Generalitat Valenciana in the program Prometeo and the EU Integrated Infrastructure Initiative Hadron Physics Project  under Grant Agreement n.227431.

%
%

\end{document}